# With no Color and Scent (part II): Metal and Alloy Microstructures—Handmade Replicas of Natural Objects

G.K. Strukova[1], G.V. Strukov[1], A.Yu. Rusanov[2], S.V. Egorov[1]

[1] Institute of Solid State Physics RAS, Academician Osipyan Str.2, 142432, Chernogolovka, Russia
[2] Applied radiophysics LTD, Severniy, 1, Chernogolovka, Russia

As a continuation of work on metal and alloy "plants" synthesis on porous membranes by means of pulsed current electroplating volume metallic microstructures resembling such natural objects as shells, cabbage leaves, mushrooms are grown and presented in their modest elegance. Such structures are formed from PdNi and PdCo alloys as well as Ag, Cu and Ni in conditions defined by the shape of membrane pores and the parameters of the pulsed current. It is shown that the obtained complex structures are formed by layers of metallic nanowires as a result of their self-assembly while growing during the pulsed current electroplating process. Depending on the shape of the membrane and the regime of the pulsed current electroplating either one type of shell-like structures or various structures can be grown.

## 1. Introduction

The growth and morphogenesis mechanisms of biological objects have always been attracting scientists' attention [1-3].

In our previous work [4] various plant-like structures of submicron size with nanosized Ag, PdNi, PbIn elements were grown on porous membranes by means of computer controlled pulsed current electroplating. The major shape formation occurred after the nanowires appeared on the surface of the membrane. For instance, various volume structures of PdNi alloy were grown by pulsed current electroplating with different parameters [4]. A suggestion that the reason for such resemblance of the obtained metallic nano- and microstructures with certain types of plants are common laws of material growth on porous membranes has been made.

The aim of the present work is to demonstrate the possibility of the controlled growth of the complicated metal and alloy structures of a certain type, e.g. resembling shells or cabbage leaves. Using scanning electron microscopy we also hope to reveal more details of the mechanism behind the growth of obtained complex volume structures.

## 2. The experiment. Growing metallic structures

In order to grow metallic microstructures electrolyte aqueous solutions were used.

Electrolyte for PdNi alloy electroplating contained (g/l):  $PdCl_2$ -6,0; $NiCl_2 \cdot 6H_2O$ -130,0; $NH_4Cl$ -75, ammonium sulphamate $NH_4SO_3NH_2$ -100,0; $NaNO_2$ -30; aqueous solution of $NH_3$ – up to pH=8; nickel-plating electrolyte contained (g/l): $NiSO_4 \cdot 7H_2O$ -220, $NiCl_2 \cdot 6H_2O$ -60, $H_3BO_3$-35; Ag-plating electrolyte contained (g/l): $AgNO_3$ -40,0; sulphosalicylic acid $C_7H_6O_6S \cdot 2H_2O$ -105; $(NH_4)_2CO_3$ -25; $(NH_4)_2SO_4$ -70; copper-plating electrolyte contained (g/l): $Cu(BF_4)_2$ -250; $H_3BO_3$-30; $HBF_4$ -15. PdNi electroplating was performed at 35 – 50 °C, while other metal and alloys were grown at 20-25 °C. Various PdNi structures were grown on porous AlOx membranes with 100 nm pores. For growing shell-like structures polymer membranes were used with 50-100 nm pores.

The procedure of the metal and alloy structure pulsed current electroplating was similar to the one, described in [4]. The number of current pulses, their amplitude and length as well as the duration of pauses between pulses were computer controlled. The pulse frequency was in the range of 30-166 Hz, while the duty cycle was varied between 50 and 99 %. For obtaining the assortment of PdNi alloy structures on oxide membrane a number **n** of current pulses of fixed frequency and duty cycle was sent through the solution. In order to grow specifically shell-like structures on polymer membrane **n** series of current pulses, each combined of **n$_1$** pulses with **I$_1$, ν$_1$, d.c.$_1$** and **n$_2$** pulses with **I$_2$, ν$_2$, d.c.$_2$** respectively, were performed sequentially, see **Table 1** for details.

**Table 1.** Pulsed current metal and alloy structures electroplating details.

**I**- current pulse amplitude, **n-** number of current pulses, **v-** pulse frequency, **d.c.-** current pulse duty cycle, **d.c.**=100 $T_{on}/(T_{on} + T_{off})$%, where $T_{on}$- current pulse length, a $T_{off}$ – pause between following pulses.

| Material | Membrane | Pore diameter, nm | $I_1$, $I_2$, mA | $n_1$, $n_2$, n | $v_1$, $v_2$, Hz | $d.c._1$, $d.c._2$, % | Fig. |
|---|---|---|---|---|---|---|---|
| PdNi | Oxide | 100 | 60 | 4000 | 30 | 90,9 | 1a,b |
| PdNi | Polymer | 50-100 | 100 50 | 200-1000 100-500 30-100 | 166 43,5 | 50 87 | 3a,b,c,d,f |
| Ag | Polymer | 50-100 | 30 100 | 500 200 40-60 | 166 43,5 | 50 87 | 3e |
| Ni | Polymer | 50-100 | 200 100 | 100-500 50-200 | 166 43,5 | 50 87 | 9a,b,c |

Scanning electron microscope SUPRA-50 VP was employed for taking the images of the nano- and submicron structures grown on the surface of membranes.

## 3. Results and discussion

Our first work [4] contained images of various volume PdNi alloy structures formed by nanowires, the growth of which was continued after their reaching the surface of the porous oxide membrane.

Among previously described "algae", "flower cabbage", "patty pan squashes" we have also payed attention to bunches of wires that grow from separate points (Fig. 1), branch and broaden upwards forming broom-like and convex-concave structures looking like "shells".

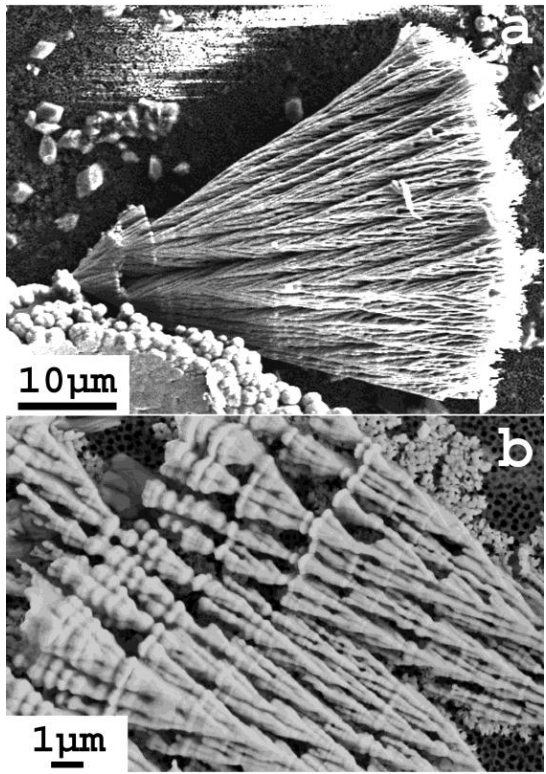

**Fig. 1.** Bunches of nanowires from PdNi alloy ("broom") growing from separate points.

Since the pulse generator maintains the programmed current intensity, the variety of obtained structures is provided, it seems, by the different forms of pores and their placing in the membrane. Indeed the membranes used had significantly different pore sizes and forms. We have succeeded in preparing pores of definite configuration on a polymer membrane, which together with precisely fixed regime of pulsed current electroplating, results in single type convex-concave structures resembling shells.

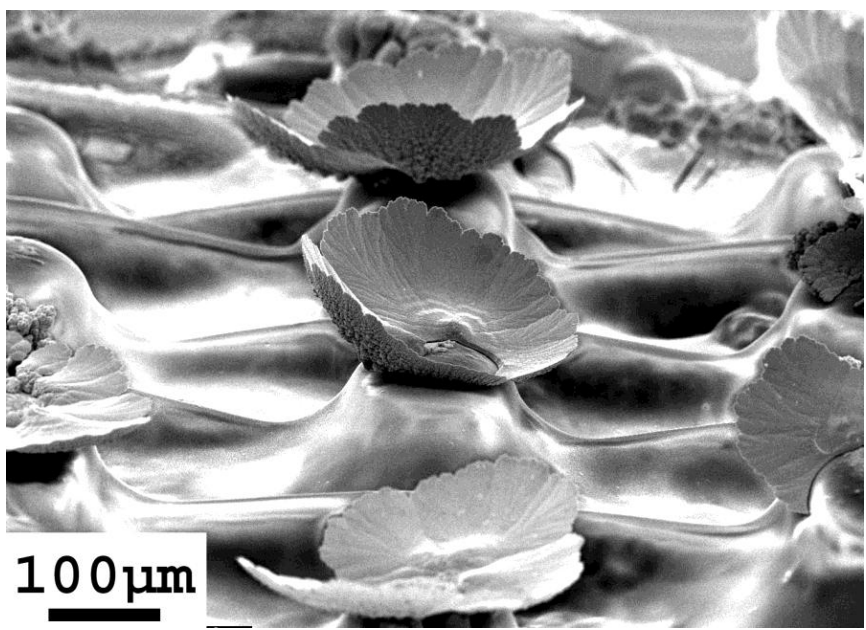

**Fig. 2.** Rows of "shells" grown from specifically prepared pores.

Differently programmed electroplating process allows growing certain variants of "shells": openwork opened and half opened shells, blooming flower, cup, brimmed hat or cabbage leaves.

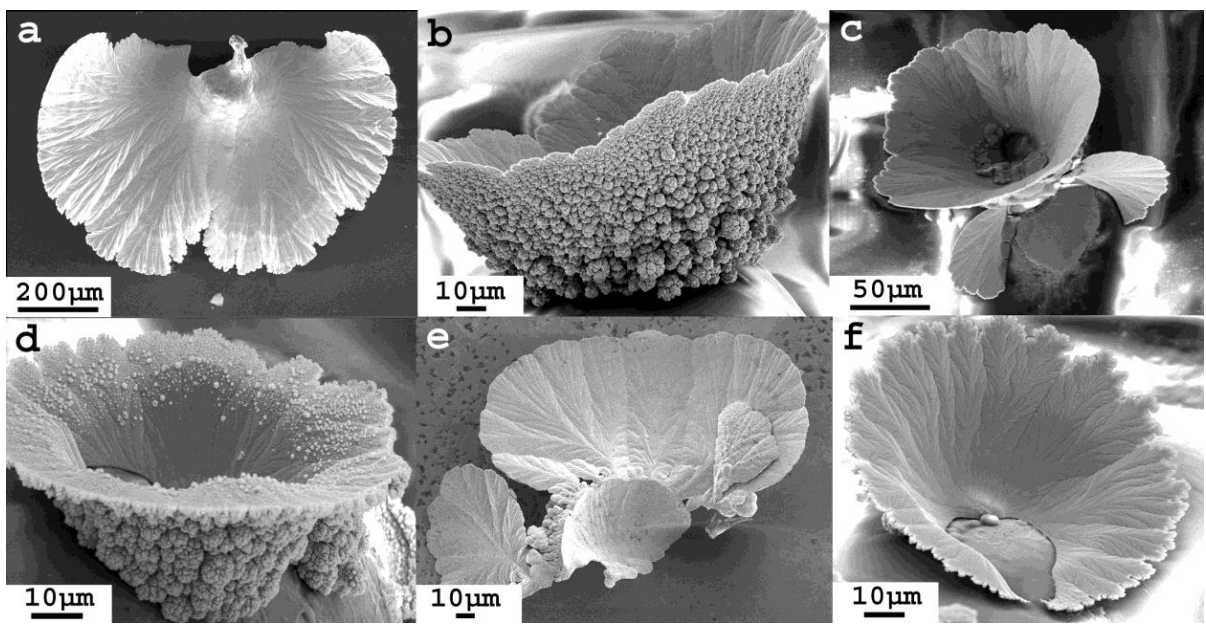

**Fig. 3.** Types of shells: a-"bird", b-"bowl", c-"flower", d-"hat", e-"cabbage leaves", f-"shell with a pearl".

The obtained structures have several characteristic features. The growth of the shell sides starts from its root. "The root" or "the bottom" of shell is a round or oval area generated by the beams of layered wires and can be compared to a stump. The edge of this area is a circle, oval or unclosed horseshoe-like line. Bundles of nanowires growing from separate points on this line form the walls of shells. These bunches broaden upwards by means of branching. Such brooms, growing from separate points, were revealed among various volume PdNi alloy structures shown in Fig. 1-3.

Fig. 4 demonstrates the shell edges where it is visible that the walls are formed by several twisted nanowires.

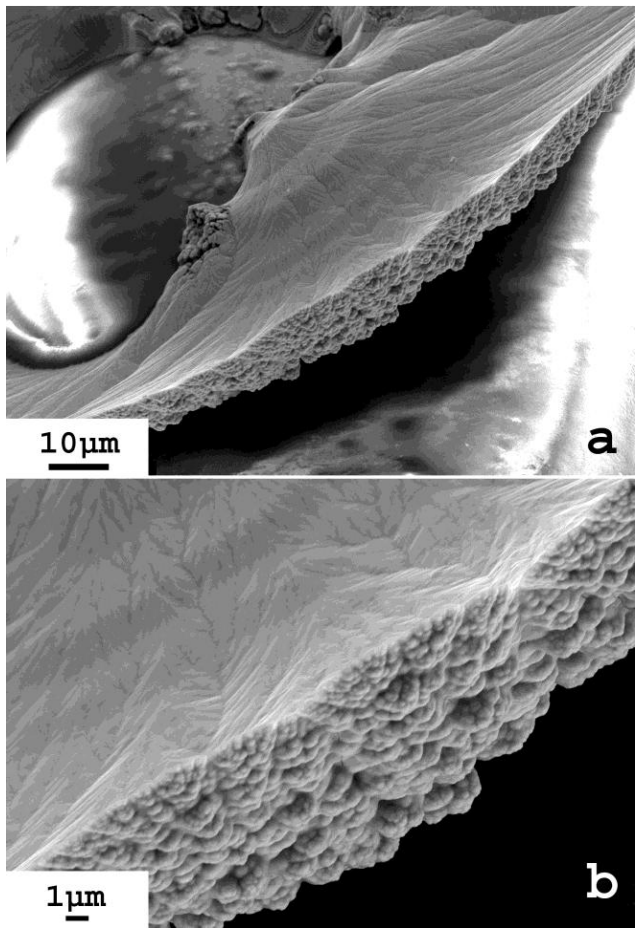

**Fig. 4.** Layered structure of the shell.

Fig. 5a shows the inner concave side of the shell, Fig. 5b shows outward side of it .

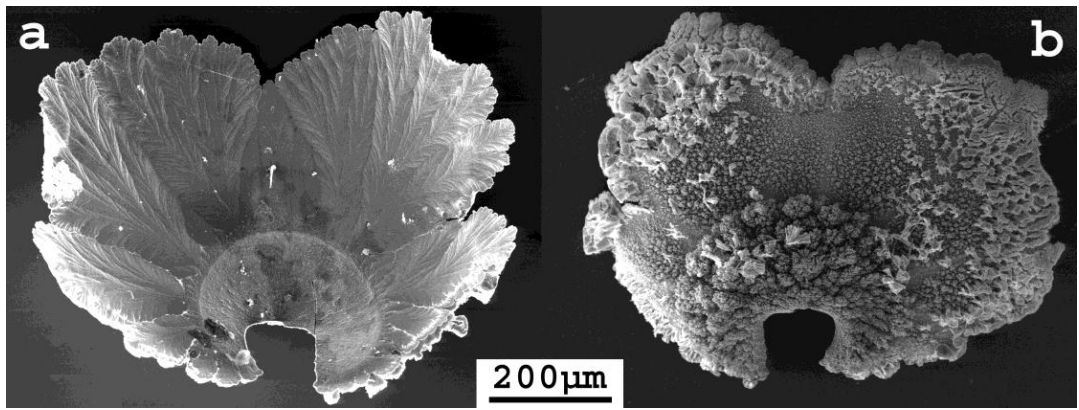

**Fig.5.** Inner and outer sides of the "shell".

A relatively smooth inner concave side of the shell with a visible pattern of intertwined bunches of wires, is a result of self-assembling process (see also Fig. 2-4). The outward side of the shell can be decorated with "flowers" formed by ends of nanowires with "buds" (see also Fig. 3 b,d), covered with fringe or bristle of branching fiber (Fig. 6).

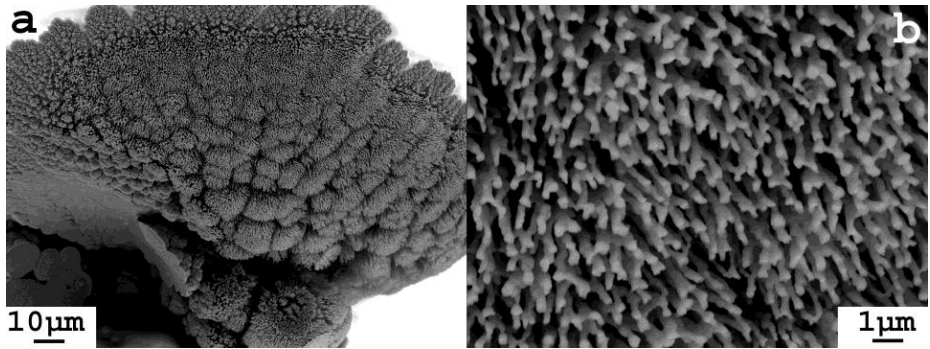

**Fig.6.** Outer side of the shell: bristle of branching fiber.

Fig. 7 demonstrates the bunches of nanowires growing anew from the "flowers" on the outer side. It can be seen that closer towards the edge, the bunches of wires are more densely entwined, one-way oriented and forming a dense bristle. There are recurrent thickenings of wires and their diameter is 70-100 nm (Fig. 7d).

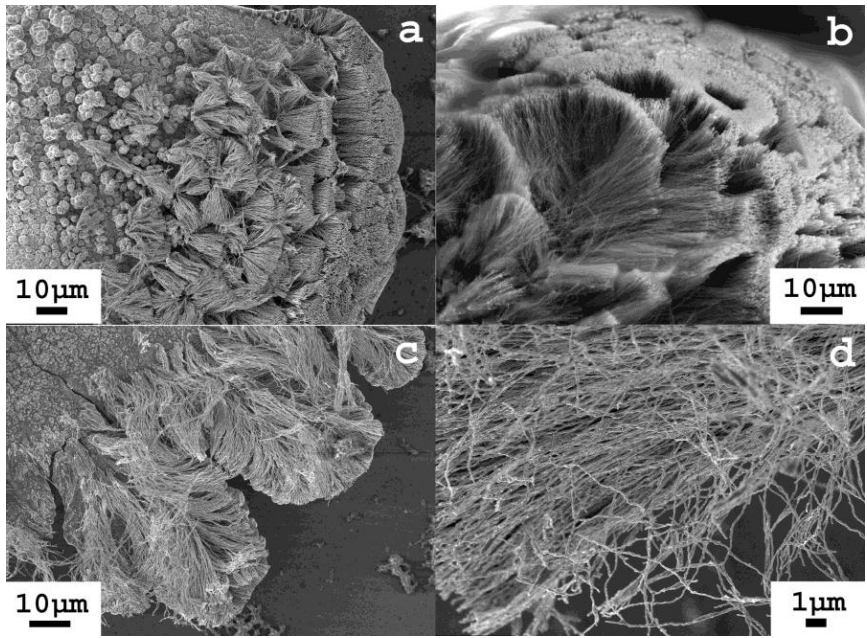

**Fig.7.** The outward side of the shell: a,b,c – the bunches of nanowires; d – nanowires.

Apart from the layer-lines of the branches that spread from the "root" to the wall of the shell ("meridians",- it can be seen in all figures), visible on the inner side, there are also transverse lines ("parallels") that obviously correspond to the amount of current pulses series n (it can be seen in Fig. 8a, 8b). The further from the shell "root" the lesser is the distance between the parallels.

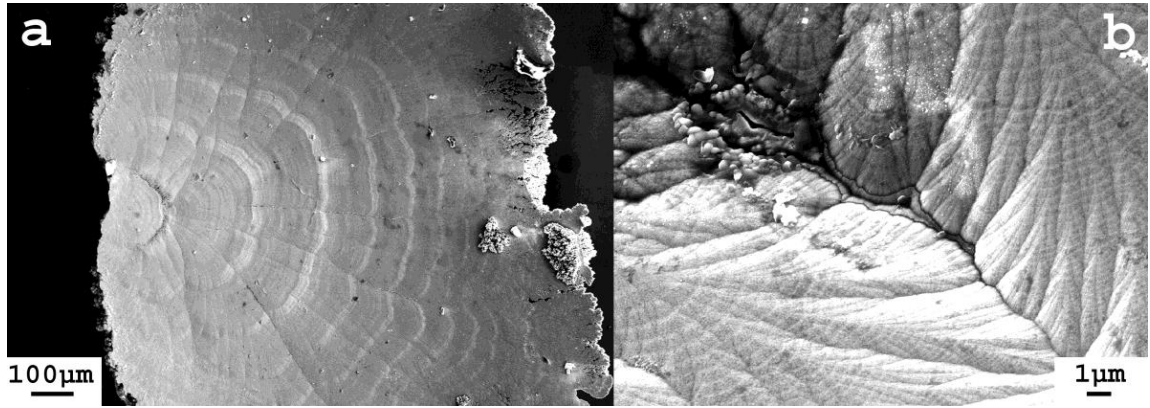

**Fig.8**. The inner side of shell: "meridians" and "parallels".

When using nickel-plating electrolyte under experimental conditions for growing shells different sphere-like structures, single (Fig. 9b) and consisting of several spherical elements (Fig. 9a) as a result of fractal branching were obtained.

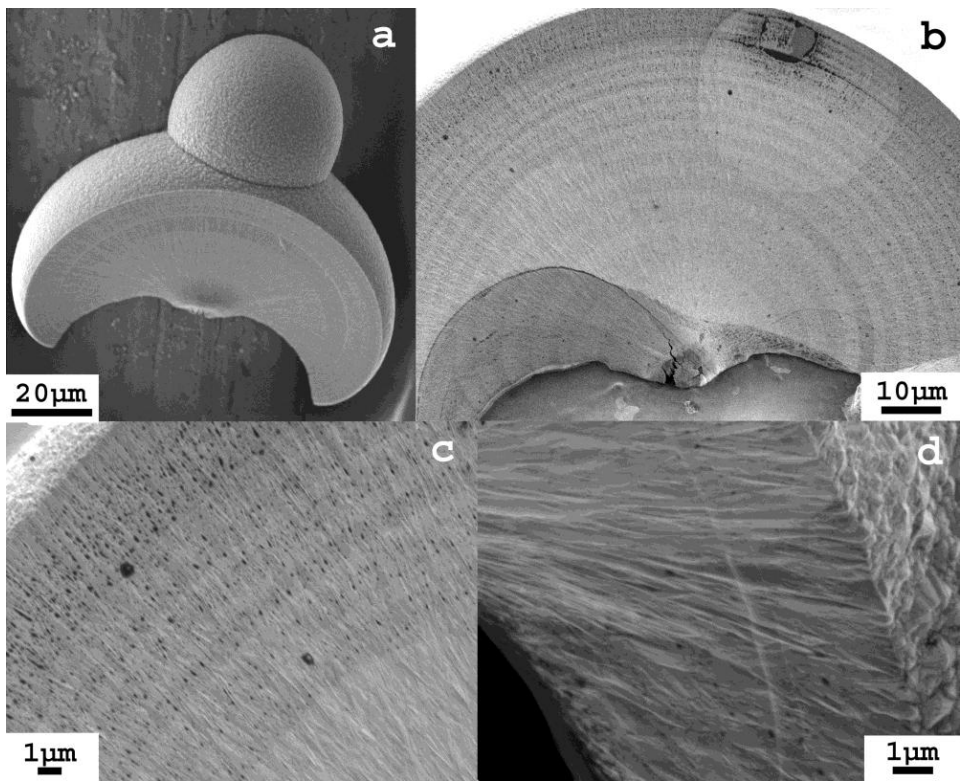

**Fig. 9.** "Mushrooms" of nickel.

It is certainly obvious that similar to the shell-like structures these, spheric structures resembling mushrooms are also growing out of a "root" (Fig.9 a,b). The inside of the spheroids displays patterns formed by self-assembling arranging of nanowires' bunches ("meridians" - Fig.9 c,d) and lines corresponding to the changes of current pulses series ("parallels"- Fig.9 a,b,c) where dark spots of holes in nanowires is visible (Fig.9c). It is visible that from the growth starting point (the "root") towards the edge the distance between parallels decreases and the amount and size of the holes increases, that is the layers of nanowires become less dense. The outward surface of the spherical mushroom is formed by nickel submicron crystallites. Their morphology corresponds to the typical morphology of nickel (211) oriented crystallites that can be obtained from this electrolyte with pulsed current electroplating at an ordinary metal substrate.

## 4. Conclusion

Elegant spatial PdNi alloy structures ranging in size from several microns to hundreds, resembling natural objects such as shells, cabbage leaves, mushrooms were obtained on porous membranes using method of pulsed current electroplating. Specific conditions — combination of membrane and pulsed current regime — that guarantee growing "shells" were found.

The "shells" were grown of Cu and PdCo alloy. Similar structures—cabbage leaves—were grown of silver. Spheroidal nickel structures resembling mushrooms were grown under similar conditions.

The obtained objects, replicating the natural ones, are formed due to the self-assembling arranging of nanowires during the pulsed current electroplating process.

Depending on the membrane geometry and the pulsed current regime, either structures of one certain type (shells) or a variety of structures can be grown of the PdNi alloy. The same results apply to Ag. Thus it characterizes the discovered method as a general method of morphogenesis of ordered volume metal and alloy nano- and mictrostructures.

The resemblance of the obtained structures shape to their natural counterparts (plants, shells, mushrooms) makes it possible to suspect the analogy between the mechanisms of growth.

The common features are the growing from porous membranes of nanosized fibres, further formation of various volume structures by joining of nanoclusters in "growth points", spatially oriented branching and self-assembling arrangement of nanowires during the growing process.

It can also be supposed that the pulsed growing regime plays a significant role in the shape formation process.

## 5. Acknowledgments

The authors thank colleagues from the Laboratory for Superconductivity: Anna Rossolenko and Ivan Veshchunov for their great help in this work.

**Table of contents entry**

Keywords: pulsed current electroplating in porous membranes, metallic microstructures, self-assembly

As a continuation of work on metal and alloy "plants" synthesis on porous membranes by means of pulsed current electroplating volume metallic microstructures resembling such natural objects as shells, cabbage leaves, mushrooms are grown and presented in their modest elegance. Such structures are formed from PdNi and PdCo alloys as well as Ag, Cu and Ni in conditions defined by the shape of membrane pores and the parameters of the pulsed current. It is shown that the obtained complex structures are formed by layers of metallic nanowires as a result of their self-assembly while growing during the pulsed current electroplating process. Depending on the form of the membrane and the regime of the pulsed current electroplating either one type of structures ("shells") or various structures can be grown.

## With no Color and Scent (Part II): Metal and Alloy Microstructures—handmade replicas of natural objects

**G.K. Strukova, G.V. Strukov, A.Yu. Rusanov, S.V. Egorov**

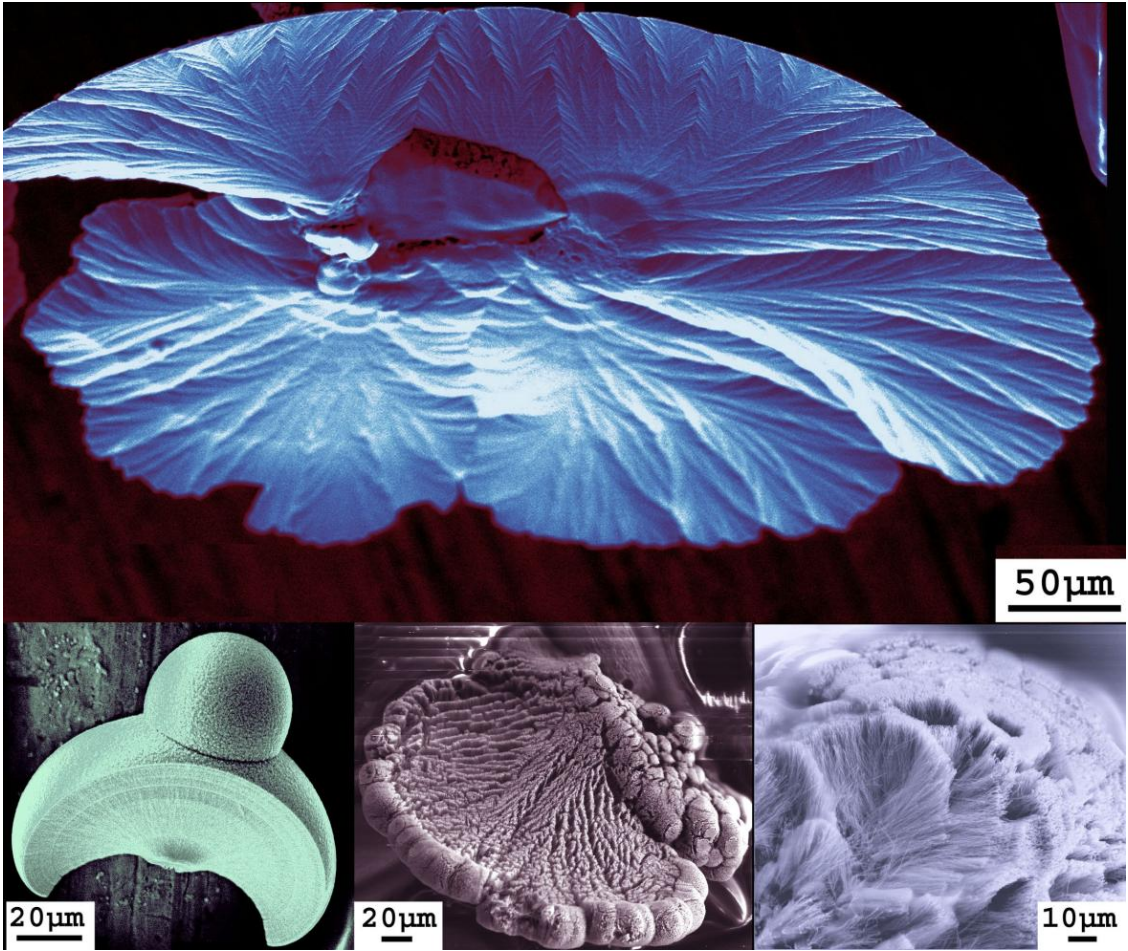

**Fig.10ad**